\documentclass[aps,twocolumn,prc,showpacs]{revtex4}
\usepackage{mathrsfs}
\usepackage{longtable,lscape}
\usepackage{txfonts}
\usepackage{amssymb}
\usepackage{indentfirst}
\usepackage{graphicx,,booktabs}
\usepackage{color}
\usepackage{amssymb}
\usepackage{epsfig}

\newcommand{\vsig}{\mbox{\boldmath$\sigma$\unboldmath}}
\newcommand{\vrho}{\mbox{\boldmath$\rho$\unboldmath}}
\newcommand{\vlamb}{\mbox{\boldmath$\lambda$\unboldmath}}
\newcommand{\veps}{\mbox{\boldmath$\epsilon$\unboldmath}}

\newcommand{\be}{\begin{equation}}
\newcommand{\ee}{\end{equation}}
\newcommand{\bea}{\begin{eqnarray}}
\newcommand{\eea}{\end{eqnarray}}
\newcommand{\bean}{\begin{eqnarray*}}
\newcommand{\eean}{\end{eqnarray*}}

\newcommand{\gapproxeq}{\lower
.7ex\hbox{$\;\stackrel{\textstyle >}{\sim}\;$}}
\newcommand{\lapproxeq}{\lower
.7ex\hbox{$\;\stackrel{\textstyle <}{\sim}\;$}}

\begin{document}

\title{Understanding the newly observed $\Omega_c$ states through their decays}
\author{
Kai-Lei Wang$^{1}$, Li-Ye Xiao$^{1}$, Xian-Hui Zhong$^{1,3}$~\footnote {E-mail: zhongxh@hunnu.edu.cn}, Qiang Zhao$^{2,3,4}$~\footnote {E-mail: zhaoq@ihep.ac.cn}}

\affiliation{ 1) Department
of Physics, Hunan Normal University, and Key Laboratory of
Low-Dimensional Quantum Structures and Quantum Control of Ministry
of Education, Changsha 410081, China }

\affiliation{ 2) Institute of High Energy Physics and Theoretical Physics Center for Science Facilities,
Chinese Academy of Sciences, Beijing 100049, China}

\affiliation{ 3) Synergetic Innovation Center for Quantum Effects and Applications (SICQEA),
Hunan Normal University, Changsha 410081, China}
\affiliation{ 4)  School of Physical Sciences, University of Chinese Academy of Sciences, Beijing 100049, China}


\begin{abstract}

The strong and radiative decay properties of the low-lying $\Omega_c$ states are studied in a constituent quark model. We find that the newly observed $\Omega_c$ states by the LHCb Collaboration can fit in well the decay patterns. Thus, their spin-parity can be possibly assigned as the following: (i) The $\Omega_c(3000)$ has $J^P=1/2^-$ and corresponds to the narrow $1P$ mixed state
$|1^2P_{\lambda}\frac{1}{2}^-\rangle_1$, its partner $|1^2P_{\lambda}\frac{1}{2}^-\rangle_2$
should be a broad state with a width of $\sim 100$ MeV.
(ii) The $\Omega_c(3050)$ and $\Omega_c(3066)$ can be assigned to be two $J^P=3/2^-$ states, $|1^4P_{\lambda}\frac{3}{2}^-\rangle$ and $|1^2P_{\lambda}\frac{3}{2}^-\rangle$, respectively.
(iii) The $\Omega_c(3090)$ can be assigned as the $|1^4P_{\lambda}\frac{5}{2}^-\rangle$ state with $J^P=5/2^-$.
(iv) The $\Omega_c(3119)$ might correspond to one of the two $2S$ states of the first radial
excitations, i.e. $|2^2S_{\lambda\lambda}\frac{1}{2}^+\rangle$ or $|2^4S_{\lambda\lambda}\frac{3}{2}^+\rangle$.
\end{abstract}


\maketitle

\section{Introduction}\label{intr}

Although the existence of $\Omega_c$ states have been predicted by the quark
model for a long time, experimental information about the $\Omega_c$ spectrum has
been extremely limited during the past several decades.
The status about the $\Omega_c$ spectrum can be found in the recent reviews
~\cite{Chen:2016spr,Klempt:2009pi,Crede:2013sze,Cheng:2015iom}.
Very recently, five new narrow $\Omega_c$ states, $\Omega_c(3000)$, $\Omega_c(3050)$,
$\Omega_c(3066)$, $\Omega_c(3090)$ and $\Omega_c(3119)$, were observed
in the $\Xi_c^{+}K^-$ channel by the LHCb
Collaboration~\cite{Aaij:2017nav}. This observation can be regarded as a
significant progress towards a better understanding of the $\Omega_c$ spectrum
and immediately attracts a lot of attention from the hadron physics community.
Together with the two established ground states,
$\Omega_c(2695)1/2^+$ and $\Omega_c(2770)3/2^+$~\cite{Olive:2016xmw},
the $\Omega_c$ spectrum, for the first time, allows a more quantitative
analysis of the internal structures, quantum numbers,
and decay modes for higher excited states.

\begin{figure}[ht]
\centering \epsfxsize=8 cm \epsfbox{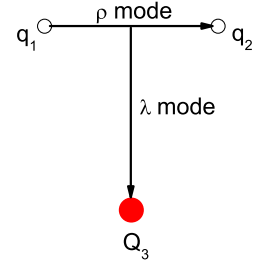}\vspace{-1.2cm} \caption{ (Color
online) $ssc$ system with $\vlamb$- or $\vrho$-mode excitations. $\vrho$ and $\vlamb$ are the Jacobi coordinates defined as $\vrho=\frac{1}{\sqrt{2}}(\mathbf{r}_1-\mathbf{r}_2)$ and $\vlamb=\frac{1}{\sqrt{6}}(\mathbf{r}_1+\mathbf{r}_2-2\mathbf{r}_3)$. $q_1$ and $q_2$ stand for the light $s$ quarks,
and $Q_3$ stands for the heavy $c$ quark.  }\label{fjcob}
\end{figure}

These newly observed $\Omega_c$ states are good candidates for
the low-lying $\Omega_c$ resonances. Since the $\Omega_c$
contains a heavy $c$ quark and two relatively light $s$ quarks,
the low-lying internal excitations will favor excitations of the
so-called ``$\lambda$-mode" in one orbital excitation in a Jacobi
coordinate between the light quarks and the heavy $c$ quark. Such a
structure is illustrated in Fig.~\ref{fjcob}.
According to the mass spectrum from various theoretical studies
~\cite{Ebert:2011kk,Maltman:1980er,Roberts:2007ni,Valcarce:2008dr,Ebert:2007nw,Garcilazo:2007eh,
Shah:2016nxi,Bali:2015lka,Yoshida:2015tia,Chen:2016phw,Chen:2015kpa,Wang:2017goq},
these newly observed $\Omega_c$ states can be organized into the first
orbital excitations ($1P$ states with $J^P=1/2^-, \ 3/2^-, \ 5/2^-$)
and the first radial excitations ($2S$ states with $J^P=1/2^+, \ 3/2^+$)
of the $\lambda$ mode, which have been summarized in
Table~\ref{wfS}. Stimulated by the newly observed $\Omega_c$ states from LHCb,
some groups have discussed their nature and possible quantum
numbers~\cite{Agaev:2017jyt,Chen:2017sci,Karliner:2017kfm,Padmanath:2017lng,Agaev:2017lip,Kim:2017jpx,
Aliev:2017led,Chen:2017gnu,Wang:2017zjw,Huang:2017dwn,Cheng:2017ove,Zhao:2017fov,Wang:2017vnc,Yang:2017rpg}.
The possible spin-parity quantum numbers suggested in the literature are collected in Table~\ref{qqq} and
there are still different views on their properties.

\begin{center}
\begin{figure}[ht]
\centering \epsfxsize=10 cm \epsfbox{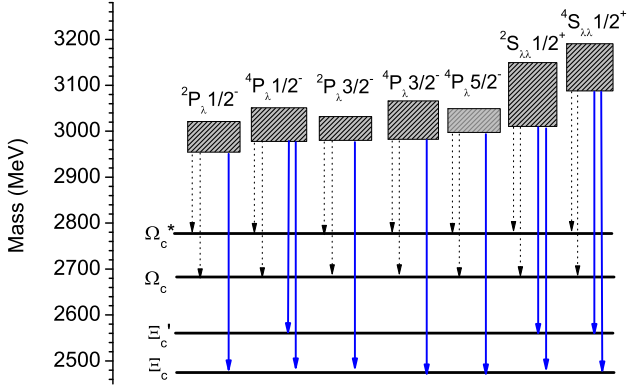}\vspace{-1.2cm} \caption{ (Color
online) Energy level of the $1P$- and $2S$-wave $\Omega_c$ states and their possible decay modes considered in present work.
The solid and dotted arrows stand for the possible decay modes with one-photon and one-kaon emissions, respectively. }\label{enle}
\end{figure}
\end{center}

It should be noted that most of these low-lying $\Omega_c$ states have masses
in the vicinity of the $\Xi_c^+ K^-$ and $\Xi_c'\bar{K}$ threshold, to which
the strong decay will almost saturate their total decay widths. Meanwhile,
for these states, their
decays will be dominated by the leading constituent quark model wavefunctions instead
of detailed structures, e.g. due to hyperfine splittings, because of their relatively small mass differences.  In other words, we
anticipate that without detailed information about the mass orderings in their
classification, one can still possibly identify the predominant feature of their
strong decay patterns for given quantum numbers. This makes it possible for
us to determine their quantum numbers based on the present available experimental
information on the partial and total widths. In addition to the hadronic decay,
we also show that the electromagnetic (EM) transitions are useful for providing
further information about their internal structures. For the low-lying $\Omega_c$
hadronic and radiative decays, the allowed decay channels are plotted
in Fig.~\ref{enle} as an illustration.

As follows, in Sec. II we first give a brief introduction to the quark model description
of the strong and radiative decay of the $ssc$ system. The numerical results are
presented and discussed in Sec.~\ref{cwr}. Finally, a summary is
given in Sec.\ \ref{suma}.

\begin{table*}[ht]
\caption{The spectrum of $1P$ and $2S$-wave $\Omega_c$ states in the constituent quark model.
The total wave function of a $\Omega_c$ state is denoted by $|N^{2S+1} L_{\sigma}J^P\rangle$.
The Clebsch-Gordan series for the spin and angular-momentum
addition $|N^{2S+1} L_{\sigma}J^P\rangle=
\sum_{L_z+S_z=J_z} \langle LL_z,SS_z|JJ_z \rangle ^{N}\Psi^{\sigma
}_{LL_z} \chi_{S_z}\phi_{\Omega_c}$ has been omitted. The details of the wavefunctions can be found in our previous work~\cite{Zhong:2007gp}. The unit of mass is MeV in the table.} \label{wfS}
\begin{tabular}{|c|c|c|c|c|c|c|c|c|c|c }\hline\hline
State                        &  Wave     & Predicted & predicted  &predicted& predicted & predicted& predicted &Observed  \\
$|N^{2S+1}L_{\sigma}$  $J^P\rangle$  & function  & mass~\cite{Ebert:2011kk}& mass~\cite{Maltman:1980er}&mass~\cite{Roberts:2007ni}& mass~\cite{Shah:2016nxi} &mass~\cite{Yoshida:2015tia}& mass~\cite{Bali:2015lka} &state\\
\hline
$|0 \ ^2 S$            \ \     $\frac{1}{2}^+\rangle$  & $^{0}\Psi^S_{00}\chi^\lambda_{S_z}\phi_{\Omega_c}$                   &2698 & 2745& 2718 &2695& 2731&2648(28) &$\Omega_c(2695)$\\
\hline
$|0 \ ^4 S$            \ \     $\frac{3}{2}^+\rangle$ & $^{0}\Psi^S_{00}\chi^s_{S_z}\phi_{\Omega_c}$                          &2768 & 2805& 2776& 2767 &2779&2709(32)&$\Omega_c(2770)$\\
\hline
$|1 \ ^2 P_\lambda$     \     $\frac{1}{2}^-\rangle$ & $^{1}\Psi^{\lambda}_{1L_z} \chi^{\lambda}_{S_z}\phi_{\Omega_c}$        &3055 & 3015& 2977&3011& 3030&2995(46)& \\
$|1 \ ^2 P_\lambda$     \     $\frac{3}{2}^-\rangle$ &                                                                        &3029 & 3030& 2986&2976&3033&3016(69)&$\Omega_c(3066)$?\\
\hline
$|1 \ ^4 P_\lambda$     \     $\frac{1}{2}^-\rangle$ &                                                                        &2966 &3040&2990&3028&3048&& \\
$|1 \ ^4 P_\lambda$     \     $\frac{3}{2}^-\rangle$ & $^{1}\Psi^{\lambda}_{1L_z} \chi^{s}_{S_z}\phi_{\Omega_c}$              &3054 &3065&2994&2993&3056& &$\Omega_c(3050)$?\\
$|1 \ ^4 P_\lambda$     \     $\frac{5}{2}^-\rangle$ &                                                                        &3051 &3050&3014&2947&3057&&$\Omega_c(3090)$? \\
\hline
$|2 \ ^2 S_{\lambda\lambda}$ $\frac{1}{2}^+\rangle$ & $^{2}\Psi^{\lambda\lambda}_{00} \chi^{\lambda}_{S_z}\phi_{\Omega_c}$   &3088 &3020&3152&3100&&&$\Omega_c(3119)$?\\
\hline
$|2 \ ^4 S_{\lambda\lambda}$ $\frac{3}{2}^+\rangle$ & $^{2}\Psi^{\lambda\lambda}_{00} \chi^{s}_{S_z}\phi_{\Omega_c}$         &3123 &3090&3190&3126&&&$\Omega_c(3119)$?\\
 \hline
\end{tabular}
\end{table*}

\begin{table*}[ht]
\caption{Spin-parity ($J^P$) numbers of the newly observed $\Omega_c$ states suggested in various works.} \label{qqq}
\begin{tabular}{c|ccccccccccc }\hline\hline
 State        &~\cite{Agaev:2017jyt} &~\cite{Chen:2017sci}&~\cite{Karliner:2017kfm}&~\cite{Padmanath:2017lng}&\cite{Chen:2017gnu}
         &\cite{Cheng:2017ove}&\cite{Wang:2017zjw}&\cite{Zhao:2017fov}&\cite{Agaev:2017lip}&\cite{Huang:2017dwn} &This work \\
\hline
$\Omega_c(3000)$&       &$1/2^-$        &$1/2^-$ ($3/2^-$)&$1/2^-$&$1/2^-$&$1/2^-$&$1/2^-$&$1/2^+$ or $3/2^+$ &$1/2^-$& &$1/2^-$\\
$\Omega_c(3050)$&       &$1/2^-$        &$1/2^-$ ($3/2^-$)&$1/2^-$&$5/2^-$&$3/2^-$&$1/2^-$&$5/2^+$ or $7/2^+$ &$3/2^-$& &$3/2^-$\\
$\Omega_c(3066)$&$1/2^+$&$1/2^+$ or $1/2^-$ &$3/2^-$($5/2^-$)&$3/2^-$&$3/2^-$&$5/2^-$&$3/2^-$&$3/2^-$       &$1/2^+$     & &$3/2^-$\\
$\Omega_c(3090)$&       &               &$3/2^-$ ($1/2^+$)&$3/2^-$&$1/2^-$&$1/2^+$&$3/2^-$&$5/2^-$       &$1/2^+$     & &$5/2^-$\\
$\Omega_c(3119)$&$3/2^+$&$3/2^+$        &$5/2^-$ ($3/2^+$)&$5/2^-$&$3/2^-$&$3/2^+$&$5/2^-$&$5/2^+$ or $7/2^+$ &$3/2^+$&$1/2^-$ &$1/2^+$ or $3/2^+$\\
\hline\hline
\end{tabular}
\end{table*}

\section{The model}\label{cr}

We apply the chiral quark model~\cite{Manohar:1983md} to the study of the hadronic decays
of the low-lying $\Omega_c$ states for rather empirical reasons. For instance,
it was shown in Refs.~\cite{Xiao:2014ura,Zhong:2010vq,Zhong:2008kd,
Zhong:2009sk,Liu:2012sj,Zhong:2007gp,Xiao:2013xi,Nagahiro:2016nsx}, that the hadronic decays of
heavy-light mesons and baryons can be reasonably described by treating the
light pseudoscalar mesons, i.e. $\pi$, $K$ and $\eta$, as a fundamental state
in the chiral quark model. Then, the decay patterns of those low-lying heavy-light mesons and baryons can
be described. The chiral quark model has also been broadly applied to various
processes involving light pseudoscalar meson productions~\cite{Li:1995si,Li:1995vi,Li:1998ni,Zhao:2002id,Li:1994cy,Li:1997gd,Saghai:2001yd,Zhao:2000iz,He:2008ty,He:2008uf,Xiao:2013hca,
Zhong:2008km,Zhong:2013oqa,Xiao:2016dlf,Zhong:2007fx,Xiao:2015gra,Zhong:2011ti,Zhong:2011ht}.
In this model, the low energy quark-pseudoscalar-meson
interactions in the SU(3) flavor basis are
described by the effective
Lagrangian~\cite{Li:1994cy,Li:1997gd,Zhao:2002id}
\begin{equation}\label{coup}
H_{m}=\sum_j
\frac{1}{f_m}\hat{I}_j\bar{\psi}_j\gamma^{j}_{\mu}\gamma^{j}_{5}\psi_j\partial^{\mu}\phi_m,
\end{equation}
where $\psi_j$ represents the $j$th quark field in the hadron;
$\phi_m$ is the pseudoscalar meson field, $f_m$ is the
pseudoscalar meson decay constant, and $\hat{I}_j$ is the isospin operator associated with the pseudoscalar meson.

Meanwhile, to treat the radiative decay of a hadron we apply
the constituent quark model which has been successfully applied to study
the radiative decays of $c\bar{c}$ and $b\bar{b}$ systems~\cite{Deng:2016stx,Deng:2016ktl}.
In this model, the quark-photon EM coupling at the tree level
is adopted as~\cite{Brodsky:1968ea}
\begin{eqnarray}\label{he}
H_e=-\sum_j
e_{j}\bar{\psi}_j\gamma^{j}_{\mu}A^{\mu}(\mathbf{k},\mathbf{r}_j)\psi_j,
\end{eqnarray}
where $A^{\mu}$ represents the photon field with 3-momentum $\mathbf{k}$. $e_j$ and $\mathbf{r}_j$
stand for the charge and coordinate of the constituent quark $\psi_j$, respectively.

To match the non-relativistic harmonic oscillator wave functions
adopted in our calculations, we should provide the
quark-pseudoscalar and quark-photon EM couplings in a
nonrelativistic form.
In the initial-hadron-rest system, the nonrelativistic form
of the quark-photon EM coupling can be written as
~\cite{Deng:2016stx,Deng:2016ktl,Brodsky:1968ea,Li:1997gd,Zhao:2002id,Li:1994cy}
\begin{equation}\label{he2}
H_{e}^{nr}=\sum_{j}\left[e_{j}\mathbf{r}_{j}\cdot\veps-\frac{e_{j}}{2m_{j}
}\vsig_{j}\cdot(\veps\times\hat{\mathbf{k}})\right]e^{-i\textbf{k}\cdot
\textbf{r}_j},
\end{equation}
while the nonrelativistic form of the quark-pseudoscalar-meson coupling can be written as
\begin{eqnarray}\label{ccpk}
H_{m}^{nr}=\sum_j\left[A \vsig_j \cdot \textbf{q}
+\frac{\omega_m}{2\mu_q}\vsig_j\cdot \textbf{p}_j\right]I_j
e^{-i\mathbf{q}\cdot \mathbf{r}_j},
\end{eqnarray}
where $A\equiv -(1+\frac{\omega_m}{E_f+M_f})$; $\vsig_j$ and $\textbf{p}_j$ stand for the
Pauli spin vector and internal momentum operator for the $j$th quark of the initial hadron;
$\mathbf{q}$ is three momentum of the emitted light meson;
$I_j$ is the flavor operator defined for the
transitions in the SU(3) flavor space
\cite{Li:1997gd,Zhao:2002id}; and $\mu_q$ is a reduced mass given by $1/\mu_q=1/m_j+1/m'_j$ with
$m_j$ and $m'_j$ for the masses of the $j$th quark in the initial and
final hadrons, respectively.

For a light pseudoscalar meson emission in a hadron
strong decays, the partial decay width can be calculated with~\cite{Zhong:2008kd, Zhong:2007gp}
\begin{equation}\label{dww}
\Gamma_m=\left(\frac{\delta}{f_m}\right)^2\frac{(E_f+M_f)|\textbf{q}|}{4\pi
M_i(2J_i+1)} \sum_{J_{fz},J_{iz}}|\mathcal{M}_{J_{fz},J_{iz}}|^2 ,
\end{equation}
while for a photon emission in a hadron
radiative decays, the partial decay width can be calculated with~\cite{Deng:2016stx,Deng:2016ktl}
\begin{equation}\label{dww}
\Gamma_\gamma=\frac{|\mathbf{k}|^2}{\pi}\frac{2}{2J_i+1}\frac{M_{f}}{M_{i}}\sum_{J_{fz},J_{iz}}|\mathcal{A}_{J_{fz},J_{iz}}|^2,
\end{equation}
where $\mathcal{M}_{J_{fz},J_{iz}}$ and $\mathcal{A}_{J_{fz},J_{iz}}$ correspond to
the strong and radiative transition amplitudes, respectively.
The quantum numbers $J_{iz}$ and $J_{fz}$ stand for the third components of the total
angular momenta of the initial and final heavy baryons,
respectively. $\delta$ as a global parameter accounts for the
strength of the quark-meson couplings. It has been determined in our previous study of the strong
decays of the charmed baryons and heavy-light mesons
\cite{Zhong:2007gp,Zhong:2008kd}. Here, we fix its value the same as
that in Refs.~\cite{Zhong:2008kd,Zhong:2007gp}, i.e. $\delta=0.557$.

In the calculation, the standard quark model parameters are adopted.
Namely, we set $m_s=450$ MeV, and $m_c=1480$ MeV
for the constituent quark masses. The harmonic oscillator parameter
$\alpha_{\rho}$ in the wave function $\psi^n_{lm}=R_{nl}Y_{lm}$ of
the $\rho$-mode excitation between the two $s$ quarks is taken as
$\alpha_{\rho}=0.44$ GeV, which is slightly larger than that of the $\rho$-mode excitation
between the two light nonstrange quarks ($\alpha_{\rho}=0.40$ GeV )
adopted in our previous work~\cite{Zhong:2007gp}. Another harmonic oscillator parameter
$\alpha_{\lambda}$ can be related to $\alpha_{\rho}$ with the relation
$\alpha_{\lambda}=[3m_c/(2m_s+m_c)]^{1/4}\alpha_{\rho}$~\cite{Zhong:2007gp}. The kaon decay constant is taken as $f_K=160$ MeV. The masses of the
well-established hadrons used in the calculations are adopted from
the Particle Data Group (PDG)~\cite{Olive:2016xmw}. With these parameters, the strong
decay properties of most of the heavy-light mesons and charmed
baryons can be reasonably well described~\cite{Liu:2012sj,Zhong:2007gp,Xiao:2013xi,Xiao:2014ura,Zhong:2010vq,
Zhong:2008kd,Zhong:2009sk}.

\begin{center}
\begin{figure}[ht]
\centering \epsfxsize=8 cm \epsfbox{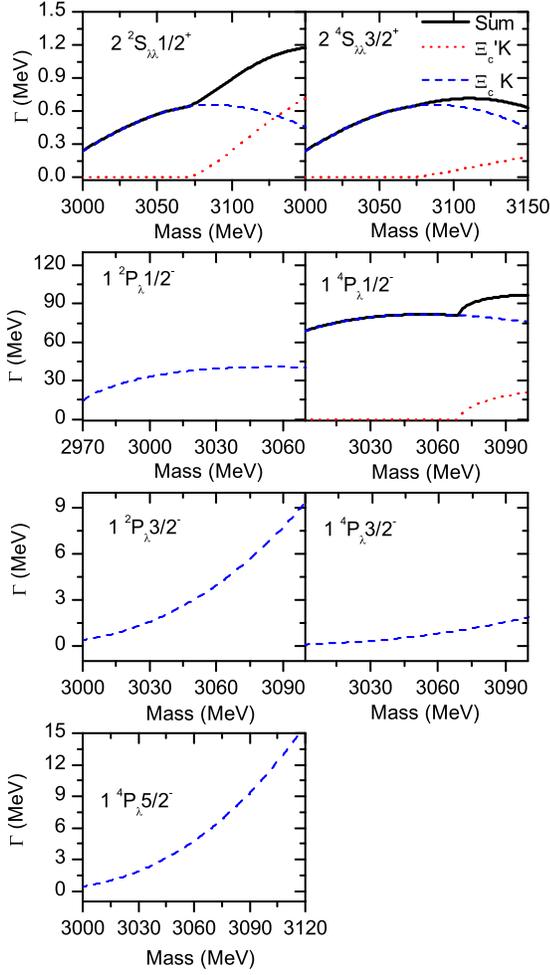}\vspace{-1.8cm} \caption{  The strong decay partial width as a function of mass. The dashed and dotted lines stand for the partial width of $\Xi_c \bar{K}$ and $\Xi_c' \bar{K}$ channels, respectively. The solid lines stand for the total width of these two channels.}\label{str}
\end{figure}
\end{center}

\begin{figure*}[ht]
\centering \epsfxsize=8 cm \epsfbox{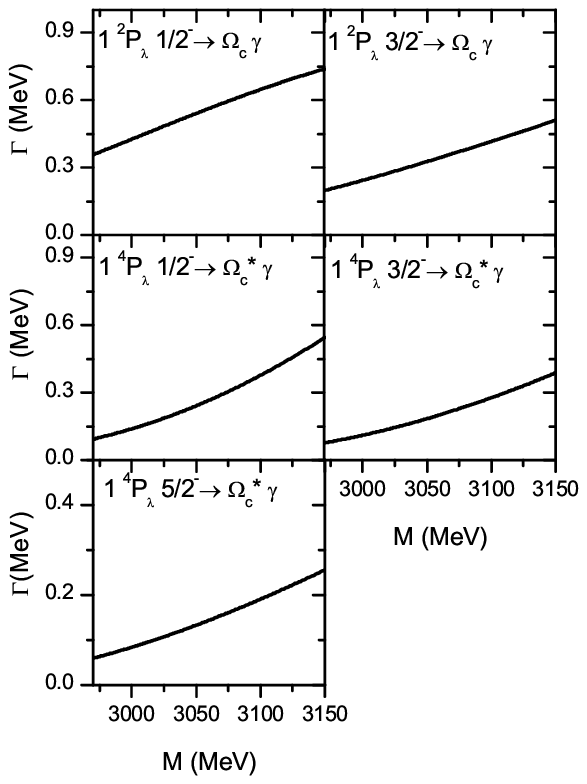} \epsfxsize=8 cm
\epsfbox{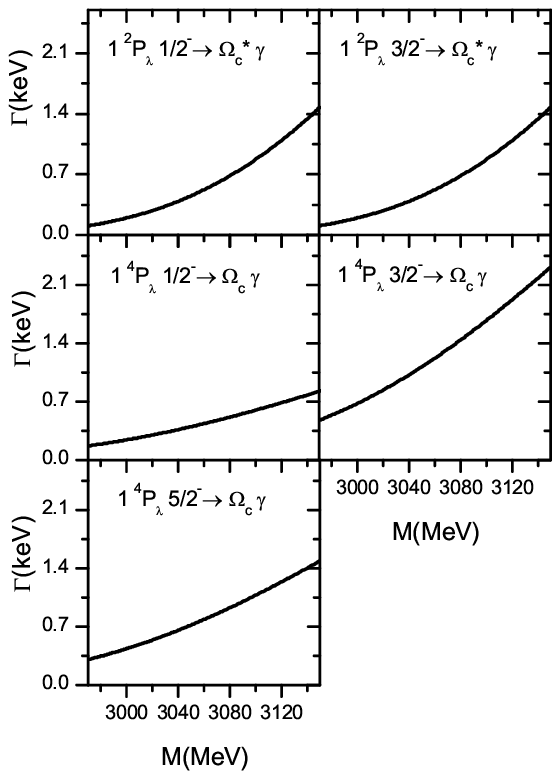} \vspace{-0.5cm} \caption{ The radiative decay partial widths of the $1P$-wave states as a function of mass. In the figure,
the $\Omega_c$ and $\Omega_c^*$ stand for the $\Omega_c(2695)$ and $\Omega_c(2770)$ states, respectively.}\label{ra}
\end{figure*}

\begin{center}
\begin{figure}[ht]
\centering \epsfxsize=9 cm \epsfbox{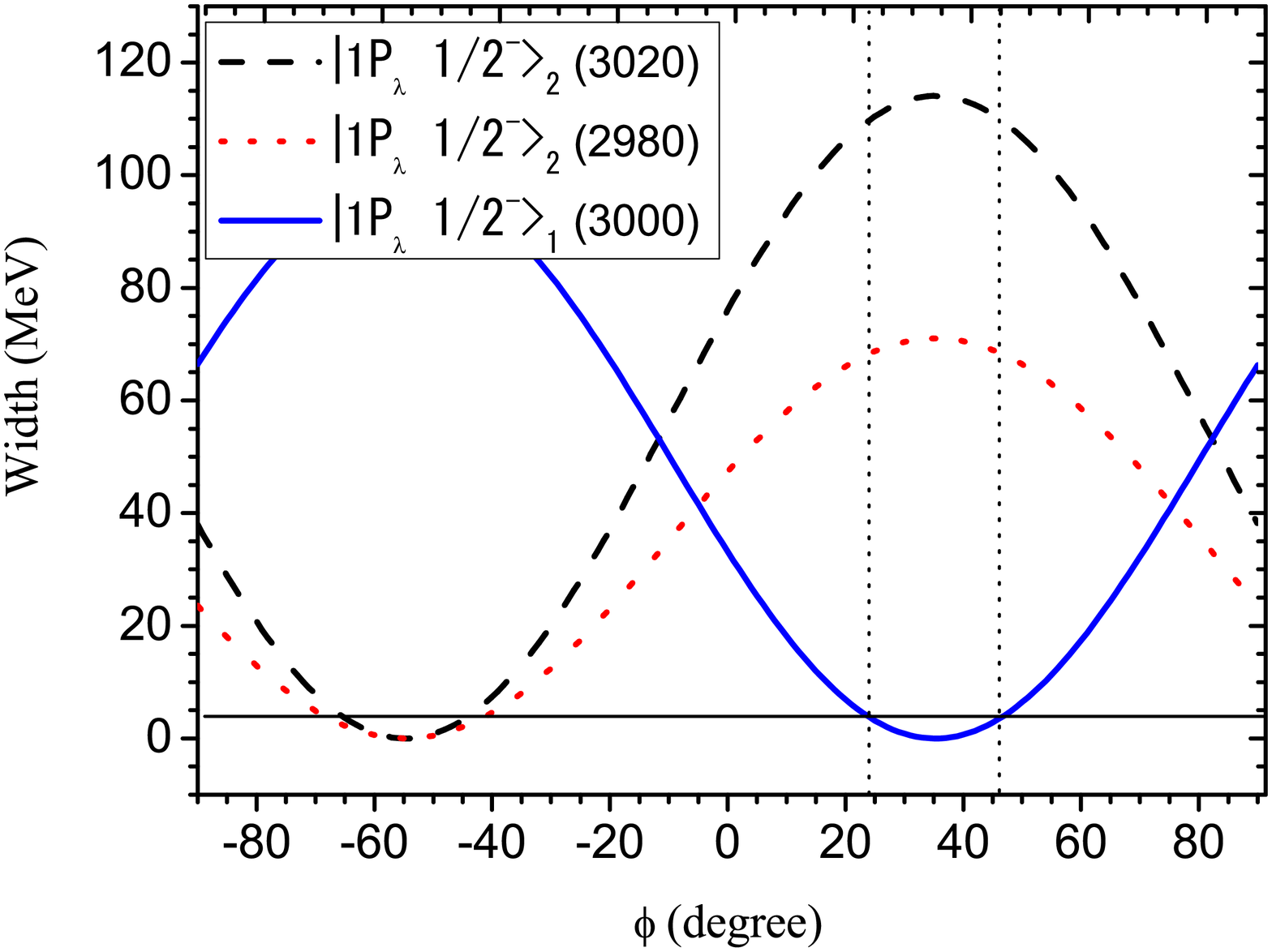}\vspace{0.0 cm} \caption{  The strong decay partial width of the $\Xi_c \bar{K}$ channel for the $J^P=1/2^-$ mixed states as a function of mixing angle $\phi$. The solid curves stand for the partial width of the mixed state $|1P_\lambda 1/2\rangle_1$ with a mass of 3000 MeV. The dashed and dotted curves stand for the partial width of the mixed state $|1P_\lambda 1/2\rangle_2$ with  masses of 3020 and 2980 MeV, respectively.  }\label{mix}
\end{figure}
\end{center}

\section{RESULTS AND DISCUSSIONS}\label{cwr}

One important feature arising from the hadronic decays of the
low-lying $\Omega_c$ states is that their hadronic decay properties are
determined by their dominant quark excitations. Within a local mass region containing several states, their decays are not sensitive to the local mass orderings which are determined by more detailed dynamics such as spin-dependent forces.
Namely, the decay pattern should not change if the mass of such
states vary within a small mass range. As a natural assumption for these
observed $\Omega_c$ states that they are most likely the $1P$ and $2S$
excited states, one can easily check that in most models the relative
partial decay widths will not change dramatically if their masses
change within 100 MeV (see figs.~\ref{str} and ~\ref{ra}). In such a sense,
the pattern arising from the relative partial widths should be more selective to their quantum
numbers instead of their masses.

In Table~\ref{tab-results} the calculations of the partial decay widths
for the $P$-wave states into $\Xi\bar{K}$, $\Xi'\bar{K}$, and radiative
decay channels are listed. It can be seen that although there are still
some uncertainties with both the experimental and theoretical results,
the magnitudes of the partial decay widths have indicated patterns
determined by the three-body quark model wavefunctions.

\subsection{$\Omega_c(3000)$}

To be more specific, the relatively low mass of $\Omega_c(3000)$ make it a good candidate for the $J^P=1/2^-$ states as the first orbital excitation states of $|1^2P_{\lambda}\frac{1}{2}^-\rangle$ or
$|1^4P_{\lambda}\frac{1}{2}^-\rangle$.
However, the quark model predicts rather broad widths for both
$|1^2P_{\lambda}\frac{1}{2}^-\rangle$ and
$|1^4P_{\lambda}\frac{1}{2}^-\rangle$ (see fig.~\ref{str}) and suggest more profound configurations with the physical state. It could happen that these two $J^P=1/2^-$ states $|1^2P_{\lambda}\frac{1}{2}^-\rangle$ and
$|1^4P_{\lambda}\frac{1}{2}^-\rangle$ can have significant mixings
for the presence of the spin-orbit interaction. Thus, we further
consider the $\Omega_c(3000)$ as a mixed state of
$|1^2P_{\lambda}\frac{1}{2}^-\rangle$ and $|1^4P_{\lambda}\frac{1}{2}^-\rangle$ by
the following mixing scheme
\begin{eqnarray}\label{3000}
\left|1 P_{\lambda}\frac{1}{2}^-\right\rangle_1&=&+\cos(\phi)\left|1^2P_{\lambda}\frac{1}{2}^-
\right\rangle+\sin(\phi)\left|1^4P_{\lambda}\frac{1}{2}^-\right\rangle,\\
\left|1 P_{\lambda}\frac{1}{2}^-\right\rangle_2&=&-\sin(\phi)\left|1^2P_{\lambda}\frac{1}{2}^-
\right\rangle+\cos(\phi)\left|1^4P_{\lambda}\frac{1}{2}^-\right\rangle,
\end{eqnarray}
where $\phi$ is the mixing angle.

Taking the $\Omega_c(3000)$ as the mixed state
$|1 P_{\lambda}\frac{1}{2}^-\rangle_1$, we plot the strong
decay width into the $\Xi_c\bar{K}$ channel as a function of the mixing angle $\phi$ in fig.~\ref{mix}.
It shows that with a mixing angle
$\phi\simeq 24^\circ$ or $47^\circ$, the measured decay width $\Gamma\simeq 4.5\pm 0.9$ MeV
of $\Omega_c(3000)$ can be well explained. Note that an intrinsic sign between $|1^2P_{\lambda}\frac{1}{2}^-\rangle$ and
$|1^4P_{\lambda}\frac{1}{2}^-\rangle$ is included which introduces the cancelation between the two transition amplitudes from these two configurations. In Ref.~\cite{Cheng:2017ove} a similar mixing mechanism for obtaining a narrow width for the $1/2^-$ state is also discussed in the basis of heavy quark spin symmetry (HQSS).

If $\Omega_c(3000)$ corresponds to the mixed state $|1 P_{\lambda}\frac{1}{2}^-\rangle_1$
indeed, the other mixed state $|1 P_{\lambda}\frac{1}{2}^-\rangle_2$
should be a broad state with a width much larger than these observed states. At this moment we do not intend to determine the mass of the broad state but only discuss its width range near the mass of $\Omega_c(3000)$. It is found that with the mass of 2980 MeV, the decay width of $|1 P_{\lambda}\frac{1}{2}^-\rangle_2$ is about $\sim 70$ MeV, while with the mass of 3020 MeV the width is about $\sim110$ MeV. This presumably suggests the difficulty of identifying it from the background in the present data sets. One notices that in the LHCb data~\cite{Aaij:2017nav} there are events excesses below $\Omega_c(3000)$ which are noted as the feed-down events from higher partially constructed $\Omega_c(X)$ states. It would be interesting to have more elaborate analysis of these event excesses to look for signals of the broad $1/2^-$. Moreover, it shows that that the lineshape of $\Omega_c(3000)$ has been distorted at the higher energy side and a broad structure is present below the narrow $\Omega_c(3000)$. further analysis of the $\Xi_c^+K^-$ invariant mass spectrum may help clarify the status of the broad partner of $\Omega_c(3000)$.

The assignment of $\Omega_c(3000)$ as the narrow $1/2^-$ state naturally leads to the dominance of the $E1$ transition in the
EM transitions of $\Omega_c(3000)\to \Omega_c(2695)\gamma$. It predicts an EM
transition partial width of $200-360$ keV, which is quite significant and can be searched for
in experiment as further evidence for its assignment. We also take the ratio between
the EM and hadronic decays as a guidance for its future studies:
\begin{equation}\label{1}
\frac{\Gamma[\Omega_c(2695) \gamma]}{\Gamma(\Xi_c\bar{K})}\simeq 5\sim 9\%.
\end{equation}
The EM transition of the state to $\Omega_c(2770)$ can also be calculated.
Taking into account the phase space factor, it predicts a rather
small partial decay width of about 10s keV, which is much
smaller than that for $\Omega_c(3000)\to \Omega_c(2695)\gamma$.

\subsection{$\Omega_c(3050)$}

The $\Omega_c(3050)$ is most likely to be the $J^P=3/2^-$ state.
It corresponds to $|1^4P_{\lambda}\frac{3}{2}^-\rangle$.
If we assign the $\Omega_c(3050)$ as $|1^4P_{\lambda}\frac{3}{2}^-\rangle$,
the two main decay channels will be $\Xi_c\bar{K}$ and $\Omega_c(2770) \gamma$
with the latter dominated by the $E1$ transition. The partial decay width
of $\Omega_c(3050)\to \Xi_c\bar{K}$ is estimated to be about 0.61 MeV and
the EM transition width of $0.33$ MeV. The EM transition of
$\Omega_c(3050)\to \Omega_c(2695)\gamma$ in the assignment of
$|1^4P_{\lambda}\frac{3}{2}^-\rangle$ will be suppressed by the $M2$
transition which leads to a small partial width of $1.12\times 10^{-3}$ MeV.
The total width reads about $0.94$ MeV which is also nearly saturated by the
hadronic and EM decays. This value is consistent with the data. The
large branching ratio of the radiative transition of $\Omega_c(3050)$
into the $\Omega_c(2770)\gamma$ channel
\begin{equation}\label{4}
Br[\Omega_c(3050)\to \Omega_c(2770) \gamma]\simeq 35\%,
\end{equation}
indicates that the radiative transition of $\Omega_c(3050)\to \Omega_c(2770) \gamma$
should be accessible in future experiment.

\subsection{$\Omega_c(3066)$}

The $\Omega_c(3066)$ can be assigned to another $J^P=3/2^-$ state,
$|^2P_{\lambda}\frac{3}{2}^-\rangle$. As the result, the $\Omega_c(3066)$
will mainly decay into
$\Xi_c\bar{K}$ and $\Omega_c(2695)\gamma$, while its decays into $\Omega_c(2770)\gamma$ will
be suppressed due to the spin-flipping $M2$ transition.
Our results have been listed in Table \ref{tab-results}.
One can see that the total width $\Gamma \simeq 4.96$ MeV
is in good agreement with the experimental data of $3.5\pm 0.4\pm 0.2$ MeV. In this scenario,
the branching ratio of the radiative transition $\Omega_c(3066)\to \Omega_c(2695)\gamma$
is predicted to be a fairly large value:
\begin{equation}\label{4}
Br[\Omega_c(3066)\to \Omega_c(2695) \gamma]\simeq 7\%.
\end{equation}
This makes the experimental measurement of the radiative decay of
$\Omega_c(3066)\to \Omega_c(2695) \gamma$ a possible way to further test its configuration.

\subsection{$\Omega_c(3090)$}

The $\Omega_c(3090)$ can be assigned to the $J^P=5/2^-$ state,
$|1^4P_{\lambda}\frac{5}{2}^-\rangle$. Its decays are governed
by the strong decay channel $\Xi_c\bar{K}$, and the predicted partial width is
\begin{eqnarray}\label{3090}
\Gamma[\Omega_c(3090)\to \Xi_c\bar{K}]\simeq \ 9.32 \ \mathrm{MeV}.
\end{eqnarray}
Following this scenario, its radiative decay
rate into the $\Omega_c(2770)\gamma$ is expected to be sizeable with a ratio of
\begin{eqnarray}\label{3090}
\frac{\Gamma[\Omega_c(3090)\to \Omega_c(2770)\gamma]}{\Gamma[\Omega_c(3090)\to \Xi_c\bar{K}]}\simeq 2\%.
\end{eqnarray}
The total width is nearly saturated by the $\Xi_c\bar{K}$ channel and with the EM transition the total
width, $\Gamma\simeq 9.5$ MeV, is in good agreement with
the measured width $8.7\pm 1.0\pm 0.8$ MeV.
To confirm the nature of $\Omega_c(3090)$, experimental measurements of the radiative decay
$\Omega_c(3090)\to \Omega_c(2770) \gamma$ are strongly recommended.

\subsection{$\Omega_c(3119)$}

The $\Omega_c(3119)$ has the highest mass among these five states but has
a narrow width of $1.1\pm 0.8\pm 0.4$ MeV. It may be assigned to be one of the first radially
excited states, i.e. either $|2^2S_{\lambda\lambda}\frac{1}{2}^+\rangle$ or
$|2^4S_{\lambda\lambda}\frac{3}{2}^+\rangle$. These $2S$ radial excitation states are found to
usually have a very narrow decay width, which is about 1 MeV (see Table~\ref{tab-2}).
Considering the $\Omega_c(3119)$ as the
$|2^2S_{\lambda\lambda}\frac{1}{2}^+\rangle$ state, we find that $\Omega_c(3119)$ should
have two main decay channels, i.e. $\Xi_c\bar{K}$ and $\Xi_c'\bar{K}$, of which the calculated partial
widths are listed in Table~\ref{tab-2}. By summing up these dominant partial
widths, the total width amounts to about 1.2 MeV, which is consistent with the
central value of the experimental data.
In this assignment, one notices that partial decay widths of the
$\Xi_c\bar{K}$ and $\Xi_c'\bar{K}$ channels are compatible.

In contrast, by assigning $\Omega_c(3119)$ as the $|2^4S_{\lambda\lambda}\frac{3}{2}^+\rangle$ state,
we find that it mainly decays into $\Xi_c\bar{K}$ and $\Xi_c'\bar{K}$ channels.
Also, the partial width into $\Xi_c\bar{K}$ will
be much larger (about a factor of 6) than into the $\Xi_c'\bar{K}$ channel. The calculated partial decay widths in
this assignment are also listed in Table~\ref{tab-2}. The measurement of the partial decay
widths into these two channels should allow a determination of the quantum number and
structure of the $\Omega_c(3119)$.

\subsection{$\Omega_c(2770)$}

As a byproduct, we also study the radiative decay process $\Omega_c(2770)\to \Omega_c(2695)\gamma$
as a test of our simple model. Our predicted partial width is
\begin{equation}\label{4}
\Gamma[\Omega_c(2770)\to \Omega_c(2695)\gamma]\simeq \ 0.89 \ \mathrm{keV},
\end{equation}
which is in good agreement with other predictions
in Refs.~\cite{Aliev:2014bma,Majethiya:2009vx,Dey:1994qi,Wang:2009cd}.
Interestingly, one notices that the lattice QCD simulation yields a rather small
value for this quantity at nearly physical pion mass~\cite{Bahtiyar:2015sga}, which is about
an order of magnitude smaller than phenomenological model calculations.
Finally, it should be mentioned that the decay widths of the
low-lying $S$ and $P$-wave charmed baryons,
such as $\Sigma_c(2455,2520)$, $\Lambda_c(2593,2625)$ and $\Xi_c(2645,2815)$, predicted within
our nonrelativistic constituent quark model~\cite{Zhong:2007gp,Liu:2012sj} are in good agreement with
the relativistic quark model predictions~\cite{Ivanov:1999bk,Lyubovitskij:2003pn}
and the experimental data~\cite{Olive:2016xmw},
which indicates that the relativistic effects are relatively small for these processes.

\begin{table*}[htb]
\begin{center}
\caption{The decay properties of the $P$-wave states compared with the observations. $\Gamma_{\mathrm{total}}^{\mathrm{th}}$ stands for the total decay width calculated in present work, while $\Gamma_{\mathrm{total}}^{\mathrm{exp}}$ stands for the total width obtained from the LHCb experiments. The unit of mass and width is MeV in the table. }\label{tab-results}
\begin{tabular}{c|ccccccccc}
\hline\hline
state    \ \ \ \                      &Mass \ \ \ \  &$\Gamma(\Xi_c\bar{K})$  \ \ \ \   & $\Gamma(\Xi_c'\bar{K})$ \ \ \ \   &$\Gamma[\Omega_c(2695)\gamma]$ \ \ \ \ &$\Gamma[\Omega_c(2770)\gamma]$  \ \ \ \  &$\Gamma_{\mathrm{total}}^{\mathrm{th}}$\ \ \ \ \ \  &  $\Gamma_{\mathrm{total}}^{\mathrm{exp}}$ & Possible assignment \\
\hline
$|1P_{\lambda}\frac{1}{2}^-\rangle_1$        &3000\ \ \     &4.0       &/           &0.36/0.20                &$0.02/0.08$       &4.38/4.28\ \ \ \ \ \       &$4.5\pm0.9$& $\Omega_c(3000)$\\ \hline
$|1^4P_{\lambda}\frac{3}{2}^-\rangle$        &3050\ \ \     &0.61       &/           & $1.12\times10^{-3}$               &0.33       &0.94\ \ \ \ \ \       &$0.8\pm0.3$&$\Omega_c(3050)$\\ \hline
$|1^2P_{\lambda}\frac{3}{2}^-\rangle$        &3066\ \ \     &4.61       &/           &  $0.35$              &$5.68\times10^{-4}$       &4.96\ \ \ \ \ \       &$3.5\pm0.4$&$\Omega_c(3066)$\\ \hline
$|1^4P_{\lambda}\frac{5}{2}^-\rangle$        &3090\ \ \     &9.32      &0.03           &$1.00\times10^{-4}$ &$0.18$                    &9.53 \ \ \ \ \ \      &$8.7\pm1.8$&$\Omega_c(3090)$\\ \hline\hline
\end{tabular}
\end{center}
\end{table*}

\begin{table*}[htb]
\begin{center}
\caption{The decay properties of $\Omega_c(3119)$ as the $2S$ states. $\Gamma_{\mathrm{total}}^{\mathrm{th}}$ stands for the total decay width calculated in present work, while $\Gamma_{\mathrm{total}}^{\mathrm{exp}}$ stands for the total width obtained from the LHCb experiments.
The unit of mass and width is MeV in the table. }\label{tab-2}
\begin{tabular}{c|ccccccccccccc}
\hline\hline
state    \ \ \ \                      &Mass \ \ \ \  &$\Gamma(\Xi_c\bar{K})$  \ \ \ \   & $\Gamma(\Xi_c'\bar{K})$ \ \ \ \   &$\Gamma[\Omega_c(2695)\gamma]$ \ \ \ \ &$\Gamma[\Omega_c(2770)\gamma]$    & $\Gamma_{\mathrm{total}}^{\mathrm{th}}$  &  $\Gamma_{\mathrm{total}}^{\mathrm{exp}}$ &  \\
\hline
$|2^2S_{\lambda\lambda}\frac{1}{2}^+\rangle$ &3119     &$0.60$  &$0.45$           &$2.9\times10^{-3}$     &$6.4\times10^{-4}$\             &1.15 &$1.1\pm0.8\pm0.4$   \\

\hline\hline
state    \ \ \ \                      &Mass \ \ \ \  &$\Gamma(\Xi_cK)$  \ \ \ \   & $\Gamma(\Xi_c'K)$ \ \ \ \   &$\Gamma[\Omega_c(2695)\gamma]$ \ \ \ \ &$\Gamma[\Omega_c(2770)\gamma]$ &  $\Gamma_{\mathrm{total}}^{\mathrm{th}}$  &  $\Gamma_{\mathrm{total}}^{\mathrm{exp}}$ &  \\
\hline
$|2^4S_{\lambda\lambda}\frac{3}{2}^+\rangle$ &3119     &$0.60$  &$0.11$           &$1.0\times10^{-3}$  &$8.1\times10^{-4}$             & 0.73 &$1.1\pm0.8\pm0.4$ \\ \hline\hline

\end{tabular}
\end{center}
\end{table*}

\section{summary}\label{suma}

In this work we have studied the strong and radiative decay properties of the
newly observed $\Omega_c$ states, i.e. $\Omega_c(3000)$, $\Omega_c(3050)$,
$\Omega_c(3066)$, $\Omega_c(3090)$ and $\Omega_c(3119)$, by
LHCb Collaboration in a constituent chiral quark model.
It shows that these low-lying states can be accommodated into the quark model with the consideration of proper internal excitations. In particular, the excitations of the $\lambda$ mode in the Jacobi coordinate (Fig.~\ref{fjcob}) will give rise to the main configurations of these observed states.

It is also found that for these low-lying states with masses
close to each other, their relative magnitudes of partial decay widths are
a more selective observable for the determination of their quantum numbers.
In contrast, the mass ordering patterns, which are determined by more detailed
dynamics, may not be an ideal quantity for classifying their quantum numbers
at the present stage.

As a conclusion of this investigation, the following
assignments seem to be favored in the quark model: (i) The $\Omega_c(3000)$ has
$J^P=1/2^-$ and corresponds to the narrow $1P$ mixed state
$|1^2P_{\lambda}\frac{1}{2}^-\rangle_1$. Its partner $|1^2P_{\lambda}\frac{1}{2}^-\rangle_2$
should be a broad state, which is worthy looking for in the future experiments. The lineshape distortion under the peak of $\Omega_c(3000)$ may be a signal for its presence.
(ii) Both $\Omega_c(3050)$ and $\Omega_c(3066)$ have
$J^P=3/2^-$ and correspond to the $1P$ states
$|1^4P_{\lambda}\frac{3}{2}^-\rangle$ and $|1^2P_{\lambda}\frac{3}{2}^-\rangle$,
respectively. The $\Omega_c(3050)$ is expected to have large radiative decay rates
into the $\Omega_c(2770)\gamma$ channel, while $\Omega_c(3066)$ has large radiative
decay rates into the $\Omega_c(2695)\gamma$ channel.
(iii) The $\Omega_c(3090)$ should correspond to the $J^P=5/2^-$ state $|1^4P_{\lambda}\frac{5}{2}^-\rangle$.
The radiative decay rate of $\Omega_c(3050)\to \Omega_c(2770)\gamma$ is is expected to be large as well.
(iv) The $\Omega_c(3119)$ may correspond to one of the first radially
excited $2S$ states, i.e. either $|2^2S_{\lambda\lambda}\frac{1}{2}^+\rangle$
or $|2^4S_{\lambda\lambda}\frac{3}{2}^+\rangle$.
The relative partial decay width fraction
$\Gamma(\Xi_c\bar{K})/\Gamma(\Xi_c'\bar{K})$ can distinguish these two different
assignments with either $J^P=1/2^+$ or $3/2^+$.

Finally, we emphasize that the EM transitions appear to be useful for determining the
quantum numbers of these $\Omega_c$ states in this analysis. Future experiments
measuring their radiative decay widths are strongly recommended.


\section*{  Acknowledgements }

We thank Y. X. Yao for checking some of the calculations. This work is supported, in part, by
the National Natural Science Foundation of China (Grant Nos. 11375061, 11425525, and 11521505), DFG and NSFC funds to
the Sino-German CRC 110 ¡°Symmetries and the Emergence of Structure in QCD¡± (NSFC Grant No. 11261130311),
and National Key Basic Research Program of China under Contract No. 2015CB856700.

\end{document}